# Algorithmic Decision Optimization Techniques for Multiple Types of Agents with Contrasting Interests

Mugurel Ionut Andreica

Politehnica University of Bucharest, Bucharest, Romania, email: mugurel.andreica@cs.pub.ro

## ABSTRACT

In this paper I present several algorithmic techniques for improving the decision process of multiple types of agents behaving in environments where their interests are in conflict. The interactions between the agents are modelled by using several types of two-player games, where the agents have identical roles and compete for the same resources, or where they have different roles, like in query-response games. The described situations have applications in modelling behavior in many types of environments, like distributed systems, learning environments, resource negotiation environments, and many others. The mentioned models are applicable in a wide range of domains, like computer science or the industrial (e.g. metallurgical), economic or financial sector.

## KEYWORDS
Contrasting Interests, Decision Optimization Techniques, Two-Player Games, Equitable Resource Allocation.

## 1. INTRODUCTION

In this paper I present several techniques for optimizing the decision process of agents which have contrasting interests. These agents perform their actions in multiple types of environments, and the interactions between them are based on various rules. These interactions are modelled by considering several types of two-player games, in which the agents have identical roles (i.e. they compete for achieving the victory in the game), or in which they have different roles (e.g. the first agent may ask several restricted types of questions to the second agent, and the second agent tries to maximize the number of questions asked by the first agent before finding the answer it seeks).

In Section 2 we introduce several game theoretic concepts which are useful in the following sections. In Section 3 we discuss several two-player games in which the players have identical roles. In Section 4 we discuss several games in which the players have different roles. In Section 5 we consider two agent pursuit games. In Section 6 we discuss three equitable resource allocation problems. In Section 7 we present related work and in Section 8 we conclude and discuss future work.

## 2. GAME THEORY CONCEPTS

We define in this section the main concepts and algorithmic techniques which will be used in the following sections. We consider that a game is played between two players (or two teams of players), which move in turns (one at a time, though not necessarily alternately). Each game has a state, which consists of all the relevant game parameters (e.g. positions of the two players or teams). We will consider an extra parameter $p$ which will be considered explicitly, representing the player which will perform the next move. Thus, the state of a game consists of a pair $(S,p)$, where $S$ is a tuple containing all the other parameters. If the game is impartial, then both players can perform the same set of moves, given a particular state of the game. The objective of every game is, of course, to achieve victory. We will consider only the following types of winning a game: *(1)* the winner is the player which performs the last move (the losing player cannot perform any valid move) ; *(2)* at the end of the game (when the game reaches a final state), a score is computed for each of the 2 players and the winner is the player with the largest score ; *(3)* a set of final states is given, for which the outcome (which player wins, or if the game ends as a draw) are known. Situation *2* also allows the game to end in a draw (equal score). Generally, the states of a game can be described by a directed acyclic graph *GS*, in which every vertex corresponds to a state $(S,p)$. For every state $(S,p)$, we know the states $(S',p')$ which can be reached by performing one move; *GS* contains a directed edge from $(S,p)$ to every such state $(S',p')$ For some of these states we know the outcome directly (*victory*, *draw*, or *defeat*, for the player $p$ whose turn is to move next). For the other states we will try to compute the outcomes in the case when both players play optimally. For every state $(S,p)$ we will compute a value *bestRes(S,p)*=the best result which can be achieved by the current player $p$ if the game is in state $(S,p)$; the results can be *victory*, *draw*, or *defeat*. For those states $(S,p)$ from which no move can be performed, the values *bestRes(S,p)* must be given (known in advance). Then, we will compute a topological sort of *GS* (since *GS* is acyclic) and we will compute the values for the states $(S,p)$ in reverse order of this sort. For every state $(S,p)$ we consider all the states *(S',p')* which can be reached from *(S,p)* by performing one move. If we have *bestRes(S',p')=defeat* (and *p'≠p*) or *bestRes(S',p')= victory* (and *p'=p*) for at least one of these states, then *bestRes(S,p)=victory*. Otherwise, if at least one of the considered states *(S',p')* has *bestRes(S',p')=draw*, then *bestRes(S,p)=draw*; otherwise, *bestRes(S,p)=defeat*.

If a score is computed for each player, then the algorithm changes follows. Every move *M* modifies the score of each player *q* (in the current state *(S,p)*) by a value *score(S,p,M,q)*. In general, in these games, every player attempts to maximize the difference between their score and the opponent's score (which is not necessarily equivalent to maximizing one's own score). Thus, for every state *(S,p)* we will compute *maxDif(S,p)*=the maximum difference between the score of the current player *p* and the opponent's score, if the game is in state *(S,p)*. For those states *(S,p)* of *GS* whose out-degree is *0*, the score which is obtained by the player *p* (and, possibly, even the one obtained by the opposite player) is given (it may be *0*, or some other value): thus, *maxDif(S,p)* is known for these states. For the other states we make use of the topological sort again. We traverse the states *(S,p)* in reverse order of the topological sort, like before. For a state *(S,p)* we consider all the moves *M(S,p,1), ..., M(S,p,r(S,p))*, leading to the states *(S'(1), p'(1)), ..., (S'(r(S,p)), p'(r(S,p)))*. We have *maxDif(S,p) = max{score(S, p, M(S, p, j), p)-score(S, p, M(S, p, j), opp(p))+(if p'(j)=p then maxDif(S'(j),p'(j)) else –maxDif(S'(j),p'(j))) |1≤j≤r(S,p)}*. We denote by *opp(p)* the opponent of player *p* (if the players are numbered with *1* and *2*, we can have *opp(p)=3-p*).

When the game is impartial and the players move alternately, we can drop the index *p* from the state pairs *(S,p)*; this is because after every move, it will always be the opponent's turn (i.e. *p'=opp(p)*, or *p'(*)=opp(p)*), and because the game is impartial, both players can perform the same moves (thus, we have *bestRes(S,p)= bestRes(S,opp(p))=bestRes(S)* and *maxDif(S,p)= maxDif(S,opp(p))=maxDif(S))*. In the score case, we will have *score(S, p, M(S, p, j), p)=score(S, opp(p), M(S, opp(p), j), opp(p))=score$_1$(S, M(S, j))* and *score(S, p, M(S, p, j), opp(p))=score(S, opp(p), M(S, opp(p), j), p)=score$_2$(S, M(S, j))*. We now consider the situation in which the two players play *K* parallel games. When a player's turn comes, it can perform a move in any of the *K* games (if the corresponding game still has any valid moves left). The rules for winning or losing are the same as in the case of a single game (e.g. the first player which cannot perform a move in any of the *K* games, loses the *combined* game, or the player whose score is larger wins). In this case, we can reduce the *K* games to a single game, as follows. We consider the graph *GSC* of the game, as follows. Let $Q_i$ be the state in the $i^{th}$ game *(1≤i≤K)*; $Q_i$ does not contain which player must move next in game *i*. Then, we set the state *S* of the combined game as *S=(Q$_1$, ..., Q$_K$)*, i.e. a tuple consisting of the individual states of each of the *K* games. For every state *(Q$_j$',p')* towards which there is a move from the state *(Q$_j$,p)* in *GS(j)* (i.e. the state graph of game *j*), we add a directed edge from *((Q$_1$, ..., Q$_K$), p)* to *((Q$_1$, ..., Q$_{j-1}$, Q$_j$', Q$_{j+1}$, ..., Q$_K$), p')* in *GSC (1≤j≤K)*. As before, if the players perform moves alternately, then the indices *p (p')* can be dropped. *GSC* has *V(GS(1))·...·V(GS(K))* states (where *V(GS(i))* is the number of states in *GS(i)*). We can use any of the algorithms mentioned before on *GSC*.

We will consider next three situations for the case when we do not use scores, which are not handled at all or are handled inefficiently by the algorithms described previously: *(1)* the graph *GS* of a game contains cycles; *(2)* the graph *GSC* of a combined game contains too many states; and *(3)* the graph *GS* of a game (not necessarily combined) contains too many states.

For case *(1)*, if *GS* contains cycles, then there is a chance that the game may never end. Thus, we will have to introduce extra rules. One possibility would be for the game to last for at most *TMAX* moves (after which, depending on the state of the game, one of the player wins, or the game ends as a draw). In this case, we construct a graph *GST* which contains vertices of the form *(Q,p,t) (0≤t≤TMAX)*, where *(Q,p)* is a state in *GS*. For every directed edge *(Q,p)->(Q',p')* from *GS*, we add the edges *(Q,p,t)->(Q',p',t+1) (0≤t≤TMAX-1)* in *GST*. Graph *GST* is a directed acyclic graph. Since we now the result for the states *(Q,p,TMAX)*, we can compute the game results for the other states, by using one of the algorithms described before. Another possibility is to decide that, if the game continues to infinity, then one of the players wins/loses automatically (or the game ends as a draw). We notice that the game continues to infinity if more than *V(GS)* moves are performed. Thus, we can set *TMAX=V(GS)+1*, after which we construct the graph *GST* as described above and run one of the previously mentioned algorithms on it; for the states *(Q,p,TMAX)* of *GST* we will set the result corresponding to the game continuing to infinity. Another possibility is to use the following iterative algorithm (inspired from [10]). We initially set *bestRes(S,p)=uninitialized* (for every state *S*). Then, we set *bestRes(Sfin,p)=victory*, *defeat* or *draw* (for all those states *Sfin* for which the result is given from the beginning). Then we proceed iteratively. At every iteration we consider all the states *(S,p)* with *bestRes(S,p)=uninitialized*. For every pair *(S,p)* we consider all the states *(S',p')* which can be reached if player *p* performs a move from *S*. If we find a state *(S',p')* with *p'=p* such that *bestRes(S',p')=victory*, or a state *(S',p')* with *p'≠p* such that *bestRes(S',p')=defeat* then we set *bestRes(S,p)=victory*. If all the considered states *(S',p')* have *bestRes(S',p')≠uninitialized* then we can compute *bestRes(S,p)* as described in one of the first algorithms from this section. At every iteration, at least one value *bestRes(S,p)* must change from *uninitialized* to *victory*, *defeat*, or *draw*. When no more such value changes occur, then we finish this stage. Afterwards, we run a similar algorithm again, considering at every iteration every state *(S,p)* with *bestRes(S,p)=uninitialized* and the player *p* would lose the game if the game continued to infinity. For each pair *(S,p)* we consider all the states *(S',p')* in which player *p* can move, and if *bestRes(S',p')=draw* for one of them, then we set *bestRes(S,p)=draw*. Like before, we stop when no more values *bestRes(S,p)* change. All the states *(S,p)* with *bestRes(S,p)=uninitialized* will be set to the values corresponding to the game continuing to infinity (player *p* wins, loses, or the game ends as a draw). The total number of iterations is $O(V(GS))$ and the time complexity per iteration is $O(V(GS)+E(GS))$ (where $E(GS)$ is the number of edges of the graph *GS*). As before, if the game is impartial and the players move alternately, then the index *p* can be dropped (because from a state *(S,p)* we always move to another state *(S',p')* with *p'≠p*). For case *(2)* we will consider only impartial games, in which the winner is the player performing the last move, the players move alternately and the state graph of the game is acyclic (or the state graphs of the parallel games are acyclic, in the case of a combined game). The Sprague-Grundy game theory [9] was developed for such cases. Let's assume that we have a combined game, composed of *K* parallel games. The state graph of each game *i* is *GS(i)*. For every state *Q* in *GS(i)* ($1≤i≤K$) we compute the value $G_i(Q)$=the Grundy number associated to the state *Q*. For the states *Q* from which no move can be performed we set $G_i(Q)=0$. For the other states *Q*, in reverse topological order, we compute $G_i(Q)$ as follows. Let $Q_1, …, Q_r$ be the states which can be reached from state *Q* by performing one move. Let $GQ=\{G_i(Q_j)|1≤j≤r\}$, i.e. the set composed of the Grundy numbers of the states $Q_1, …, Q_r$. $G_i(Q)=mex(GQ)$, where *mex(SA)* is the minimum excluded value from the set *SA* (i.e. the minimum non-negative integer number which does not belong to the set *SA*). For a state *Q* in *GS(i)*, if $G_i(Q)>0$, then the player whose turn to move from state *Q* is next has a winning strategy (considering only the game *i*); if $G_i(Q)=0$, then the player to move next from state *Q* cannot win in the game *i* if the other player plays optimally. The proof of these statements is simple. We notice that we have $bestRes_{(i)}(Q)=defeat$ every time we have $G_i(Q)=0$ and $bestRes_{(i)}(Q)=victory$, every time $G_i(Q)>0$ (we denoted by $bestRes_{(i)}$ the values *bestRes* computed only for the game *i*). Let's consider now that every game *i* is in the state $Q_i$. The Grundy number of the combined game (composed of the *K* parallel games) is $GC=G_1(Q_1)$ xor ... xor $G_K(Q_K)$. If $GC>0$ then the player which will perform the next move (from the state $(Q_1, …, Q_K)$) has a winning strategy; otherwise, if $GC=0$, then the player performing the next move from the state $(Q_1, …, Q_K)$ will lose the game if its opponent plays optimally. The consequence of this result is that there is there is no need to construct the composed state graph *GSC* (consisting of $V(GS(1))·…·V(GS(K))$ vertices). Within the game, whenever we are in a combined state $(Q_1, …, Q_K)$, we can evaluate the outcome based on the Grundy numbers of the independent states $Q_1, …, Q_K$, in the corresponding state graphs *GS(1), …, GS(K)*.

For case *(3)* the construction of the state graph *GS* is too complicated even when it is not a combined game. We consider the same restrictions as in case *(2)*. Thus, we cannot compute explicitly the values *bestRes* or the Grundy numbers. For some games, however, the Grundy numbers have some interesting properties, like periodicity. For these games we will attempt to find a pattern (a rule) for computing the Grundy number of any given state and we will use it for computing the Grundy numbers directly. These games will then be easily extended to combined games in which we won't have to construct the state graphs *GS(i)* explicitly.

In some cases (with the same restrictions as in case *(2)*) we only need to compute the outcome of the game for a given initial state of the game. Let's assume that we have a combined games consisting of *K* types of parallel games, containing $x(i)≥0$ instances of every type *i* ($1≤i≤K$). In such cases, we will compute $G_i(Q_i)$ (where $Q_i$ is a state of a game instance of type *i*) and then we compute $GG(i)=0$, if $x(i)$ is even, or $GG(i)=G_i(S_i)$, if *x* is odd, where $S_i$ is the initial state of game *i*. The Grundy number of the combined game is then $GG(1)$ xor ... xor $GG(K)$. This way, the numbers $x(i)$ can be very large, because we are only interested in their parity.

## 3. 2 AGENTS WITH IDENTICAL ROLES

### 3.1. A PATH GAME ON A TREE

We have a tree with *n* vertices. All of the vertices are initially unmarked. Two players play the following game. Player *A* chooses a vertex *v* and marks it. Then, player *B* chooses an unmarked vertex *u* which is adjacent to *v*, and marks it. The game continues, the two players taking turns alternately. At its turn, the current player chooses an unmarked vertex *u* which is adjacent to the vertex marked by the other player during the previous turn, and marks it. When one of the players cannot choose a vertex satisfying all the constraints when its turn comes, that player loses the game. We want to find out for which initial vertices *v* player *A* has a winning strategy against player *B*. We will first present a linear time algorithm for the case when the vertex *v* is fixed. We root the tree at vertex *v*, thus defining parent-son relationships between vertices. We make use of the notations from [4]. Then, we traverse the tree bottom-up (from the leaves towards the root) and, for each vertex *i*, we compute *win(i)=true*, if the player whose turn has come is allowed to choose vertex *i*, chooses it and has a winning strategy from now on (or *false*, if it doesn't have a winning strategy as a result of choosing vertex *i*); after choosing vertex *i*, the opposite player will have to choose only one of vertex *i*'s sons. If *i* is a leaf, then *win(i)=true*. For a non-leaf vertex *i*, *win(i)=true* if all the values *win(s(i,j))* ($1 \leq j \leq ns(i)$) of its *ns(i)* sons are *false* (i.e. whichever vertex the other player chooses next, it won't be able to win). If *win(v)=true*, then player *A* can choose vertex *v* at its first turn and has a winning strategy from now on. Obviously, we could run this algorithm with every tree vertex as the root, in order to check if player *A* could choose that vertex at its first turn. However, this approach would lead to an $O(n^2)$ solution. We will now show how we can maintain the linear time complexity. We will borrow ideas from the algorithmic framework for trees introduced in [5]. We will first choose an initial vertex *v* and run the algorithm described previously. During the algorithm we will also compute *ntwin(i)*=the number of sons *s(i,j)* of a vertex *i* for which *win(s(i,j))=true*. Then, we will traverse the tree from top to bottom, by using a Depth-First Search (DFS) starting at the root *v*. For every visited vertex *i*, we will compute *rwin(i)*=the value of *win(i)* if the tree was rooted at vertex *i*, and *ntrwin(i)*=the number of sons of vertex *i* with *win(i)=true* if *i* were the root of the tree. Obviously, *rwin(v)=win(v)* and *ntrwin(v)=ntwin(v)*. Let's assume that we visited a vertex *i* (and we have already computed *rwin(i)* and *ntrwin(i)*) and we now want to visit one of its sons *s(i,j)* (actually, we will recursively visit the subtree of *s(i,j)*). At first, we will compute *ntrwin'(i,j)=ntrwin(i)-(if win(s(i,j))=true then 1 else 0)*. *ntrwin'(i,j)* is the number of sons *q* of vertex *i* with *win(q)=true*, if *i* were the root of the tree and *s(i,j)* were not one of vertex *i*'s sons. Then, we will compute *ntrwin(s(i,j))=ntwin(s(i,j))+(if (ntrwin'(i,j)=0) then 1 else 0)*. If *ntrwin(s(i,j))=0* then *rwin(s(i,j))=true*; otherwise, *rwin(s(i,j))=false*. This way, in *O(n)* time, we were able to compute the values *rwin(\*)*. At its first turn, player *A* can choose any vertex *u* with *rwin(u)=true*.

## 3.2. GATHERING AN EVEN NUMBER OF OBJECTS

There is one pile consisting of *N* objects (*N* is odd). Two players perform moves alternately. When its turn comes, a player may remove from the pile any number of objects *x* between *1* and *K* (as long as there are at least *x* objects in the pile). The player keeps the objects he/she removed and adds them to the objects removed during previous moves. When the pile becomes empty, each player counts the number of objects he/she gathered from the pile during the game. The winner of the game is the player who gathered an even number of objects (since the total number of objects is odd, only one of the two players may gather an even number of objects). In this case, the Sprague-Grundy theory cannot be used, because the winner is not the player who performs the last move. Instead, we can use dynamic programming. We will compute two sets of values: *win[0,i]* and *win[1,i]*. *win[0,i]* is *1*, if the pile contains *i* objects, the winner must gather an even number of objects and the player whose turn is next has a winning strategy (and *0*, otherwise); *win[1,i]* is defined in a similar manner, except that the winner must gather an odd number of objects. We have *win[0,0]=1* and *win[1,0]=0*. For $1 \leq i \leq N$, we have:

$$\text{win}[0,i] = \begin{cases} 1, \text{if } \exists (1 \leq c \leq \min\{i, K\}) \text{ such that} \\ \quad \text{win}[((c \bmod 2) + 1 + ((i-c) \bmod 2)) \bmod 2, i-c] = 0 \\ 0, \text{otherwise} \end{cases} \quad (1)$$

$$\text{win}[1,i] = \begin{cases} 1, \text{if } \exists (1 \leq c \leq \min\{i, K\}) \text{ such that} \\ \quad \text{win}[((c \bmod 2) + ((i-c) \bmod 2)) \bmod 2, i-c] = 0 \\ 0, \text{otherwise} \end{cases} \quad (2)$$

If *win[0,N]=1*, then the first player has a winning strategy; otherwise, the second player has one. The time complexity of an algorithm implementing the equations above directly is $O(N \cdot K)$. This algorithm can be improved to $O(N)$, in the following way. We will compute the same sets of values as before, but we will maintain a structure *last[x,y,z]* ($0 \leq x,y,z \leq 1$), with the following meaning: the last value of *i* (number of objects in the pile) such that: the parity of the number of objects gathered by the winner is *x* (*0* for even, *1* for odd), *y=((the number i of objects in the pile) mod 2)* and *z=win[x,i]*. The new equations for *win[0,i]* and *win[1,i]* ($1 \leq i \leq N$) and the algorithm are given below:

$$\text{win}[0,i] = \begin{cases} 1, \text{if } (i - \text{last}[1 + (i \bmod 2), (i \bmod 2), 0]) \leq K \\ 1, \text{if } (i - \text{last}[((i-1) \bmod 2), ((i-1) \bmod 2), 0]) \leq K \\ 0, \text{otherwise} \end{cases} \quad (3)$$

$$\text{win}[1,i] = \begin{cases} 1, \text{if } (i - \text{last}[(i \bmod 2), (i \bmod 2), 0]) \leq K \\ 1, \text{if } (i - \text{last}[1 + ((i-1) \bmod 2), ((i-1) \bmod 2), 0]) \leq K \\ 0, \text{otherwise} \end{cases} \quad (4)$$

**GatherAnEvenNumberOfObjects:**
*last[x,y,z]=-∞*, **for all tuples** *(x,y,z)*
*win[0,0]=1; win[1,0]=0*
*last[0,0,win[0,0]]=0; last[1,0,win[1,0]]=0*

**for** *i=1* **to** *N* **do**
  *compute win[0,i] and win[1,i] using the equations above*
  *last[0, (i mod 2), win[0,i]]=i*
  *last[1, (i mod 2), win[1,i]]=i*

The values *win[0,N]* and *win[1,N]* (with odd *N*) present some unexpected patterns. For even *K*, we have *win[0,N]=0*, only if *(N mod (K+2)=1)*. For odd *K*, we have *win[0,N]=0*, only if *(N mod (2·K+2)=1)*. With these observations, we can determine in *O(1)* time which of the two players has a winning strategy. We should notice that, by computing the *win[0,i]* and *win[1,i]* values, we also solved the version of the game in which the winner has to gather an odd number of objects. The values of *win[1,N]* exhibit similar patterns. For odd *K*, we have *win[1,N]=0*, only if *(N mod (2·K+2)=(K+2))*. For even *K*, *win[1,N]=0*, only if *(N mod (K+2)=(K+1))*. Similar rules can be developed for *win[0,N]* and *win[1,N]* when *N* is even, but in this case both players may win the game: for even *N* and odd *K*, *win[0,N]=0* only if *(N mod (2·K+2)=(K+1))*, and *win[1,N]=0* only if *(N mod (2·K+2)=0)*; for even *N* and even *K*, *win[0,N]* is always *1*, and *win[1,N]=0* only if *(N mod (K+2)=0)*. A short version of the presented solution was given in [7].

### 3.3. GATHERING OBJECTS FROM A BOARD

We consider a linear board, on which *n* objects are placed, numbered from *1* to *n* (from left to right). Every object *i* has a value $v(i) \geq 0$. Two players perform moves alternately. At each turn, the current player gathers one of the two objects from the left or right end of the board. The game ends when all the objects were gathered. At the end, every player *p* computes its score *score(p)* as the sum of the values of the objects he/she gathered. Both players want to maximize the difference between their score and the opponent's score. Optimal strategies (considering that both players play optimally) can be computed using dynamic programming. First, we compute the prefix sums *SP(i)=v(1)+…+v(i)* (*SP(0)=0* and *SP(1≤i≤n)=SP(i-1)+v(i)*). With the prefix sums we can compute the sum *Sum(a,b)* of all the values of the objects in an interval *[a,b]* in *O(1)* time: *Sum(a,b)= SP(b)-SP(a-1)*. Then, we compute *smax(i,j)=*the maximum score that the current player may obtain if the board consists only of the objects from *i* to *j* (and we ignore the other objects). We have *smax(i,i)=v(i)*. For *i<j*, we will consider the pairs *(i,j)* in increasing order of *l=j-i*. Thus, we have: *for l=1 to n-1 do: for i=1 to n-l do: j=i+l; smax(i,j) = max{v(i)+Sum(i+1,j)-smax(i+1,j), v(j)+Sum(i,j-1)-smax(i,j-1)}*. When the number of objects is even and the purpose of the game is for one of the players to obtain a larger score than the other one, the first player always has a strategy which guarantees him/her a victory or a draw. Let *SumOdd* (*SumEven*) be the sum of the values of the objects numbered with odd (even) numbers. The first player can always play in such a way that it gathers all the odd (even) numbered objects: at every move it chooses the object with odd (even) number, leaving the opponent to choose between two objects with even (odd) numbers. Thus, it can play in order to obtain a score equal to the larger of the two sums (*SumOdd* or *SumEven*).

### 3.4. GATHERING CHARACTERS FROM A BOARD

We consider a linear board, on which *n* characters of an alphabet *A* are placed, numbered from *1* to *n* (from left to right). The character on position *i* is *c(i)*. Two players move alternately. Initially, they have an empty string *S*. At each turn, the current player can remove the character at the left or right end of the board and add it to the end of *S*. The purpose of the first player is to obtain the string *S* which is lexicographically minimum, while that of the second player is to obtain a string *S* which is lexicographically maximum. The outcome of the game can be computed by using dynamic programming. We compute *Sres(i,j)=*the resulting string, if the board consisted only of the characters on the positions from *i* to *j*. Obviously, *Sres(i,i)=c(i)*. Like in the previous sub-section, we consider the pairs *(i,j)* (with *i<j*) in increasing order of the value *j-i*. For a pair *(i,j)* we need to determine which player *p* will perform the move: *p=1* if *(i-1+n-j)* is even, and *p=2* if *(i-1+n-j)* is odd. If *p=1*, then *Sres(i,j)=min{c(i)+Sres(i+1,j), c(j)+Sres(i,j-1)}*, where we denoted by *A+B* the concatenation of the strings *A* and *B* (*c(i)* can also be considered as a one-character string). If *p=2* then *Sres(i,j)=max{ c(i)+Sres(i+1,j), c(j)+Sres(i,j-1)}*.

### 3.5. GATHERING MANY OBJECTS

Two players play the following game. Initially, they have *N* objects in a pile. The two players perform moves alternately and the game proceeds in rounds. At every move, each player *p* may take from the pile any number of objects *x* which belongs to a given set *S(p)* (*S(p)* always contains the number *1*), where *p=1* or *2* is the index of the player. Every player puts the taken objects aside. The player taking the last object gets to keep all of his objects, while the other player must put all the objects he/she took back into the pile. After a player takes the last object, a round finishes. At the next round, the player performing the first move is: *(case 1)* the one who took the last object in the previous round; *(case 2)* the one who did not take the last object in the previous round. The game ends when no more objects are put back into the pile. The winner of the game is the player who gathered the largest number of objects overall. The first move of the game is performed by player *1* and we want to know which of the two players will win (in either of the two cases), considering that both will play optimally. We will compute the values *Gmax(i,j,q,p)=*the maximum total number of objects which can be gathered by the player *p* (whose turn to move is next), knowing that this player has *i* objects put aside in this round, the opponent has *j* objects put aside in this round, and the pile contains *q* objects. *Gmax(0, 0, N, 1)* will be the answer to our problem (i.e. the largest number of objects that player *1* can gather; if this number is larger than *N/2*, then player *1* will win, if it is equal to *N/2* then the game ends as a draw; otherwise, player *1* will lose the game). If *q=0* then the previous player emptied the pile. Thus, the opponent gets to keep its *j* objects. We have *Gmax(i,j,0,p)* equal to *Gmax(0,0,i,p)* (for case *2*), or to *(i-Gmax(0,0,i,3-p))* (for case *1*). For *q≥1* we will consider every value *k* from *S(p)*, such that *k≤q*. If the current player *p* took *k* objects from the pile at its next move, then it would have *i+k* objects put aside, the opponent would have *j* objects put aside and there would be *q-k* objects left in the pile. At the next move, the opponent would move and would be able to gather, overall, *Gmax(j, i+k, q-k, 3-p)* objects. Thus, the current player *p* will be able to gather at most *(i+j+q)-Gmax(j, i+k, q-k, 3-p)* objects. Thus, *Gmax(i,j,q,p)= max{(i+j+q)-Gmax(j, i+k, q-k, 3-p) | k∈ S(p) and k≤q}*. We will compute the values *Gmax(i,j,q,p)* in increasing order of the sum *(i+j+q)*, starting from *i=j=q=0* (*Gmax(0,0,0,\*)=0*). The tuples *(i,j,q)* with the same sum *(i+j+q)* will be considered in increasing order of *q*. The time complexity of the algorithm is $O(N^3 \cdot max\{|S(1)|, |S(2)|\})$, which, in the worst case, is $O(N^4)$. When *S(1)=S(2)* we can drop the index *p* (and the reference *3-p*) from the states of the table *Gmax*, maintaining only the *3* indices *i*, *j*, and *q*.

### 3.6. PARALLEL TREBLECROSS

We consider a combined game, consisting of *K* types of simple games. We have *x(i)≥0* identical instances of every type of game *i* (*1≤i≤K*). An instance of a game *i* consists of a linear board containing *N(i)≥3* positions (numbered from *1* to *N(i)*). Some of these positions are unmarked, while the others are marked. During a simple game, the two players make moves alternately. At every move, the current player chooses one of the unmarked positions and marks it. When, as a result of a player's move, there are *3* consecutive marked positions, then that player wins the game. In the initial state of a simple game, there will not be any two consecutive marked positions and neither two marked positions separated by an unmarked position between them (because, in this game, the first player would win immediately). In a combined game, the next player to move can choose an unmarked position from any of the *K* games. The winner is the one obtaining three consecutive marked positions in any of the *K* games. We want to decide which of the two players (the first player, performing the first move, or the second player) has a winning strategy.

Let's consider a linear board in which some positions are marked and the others are unmarked and in which the current player cannot win at its next move. At its next move, no player will choose an unmarked position which is adjacent to a marked position (because then the other player would win at its next move) or which is located two positions away from it (i.e. if position *i* is marked, then the player will not choose the positions *i-1*, *i-2*, *i+1*, or *i+2* for its next move, if they are unmarked, unless it is forced to do so). Thus, we can consider all the unmarked positions which are at distance *1* or *2* from a marked position as being *lightly marked*. Then, the unmarked positions which are not lightly marked form a set of *R≥0* maximal disjoint intervals (composed of consecutive unmarked positions which are not lightly marked), separated by marked or lightly marked positions. The lengths of these intervals are *L(1)*, …, *L(R)* (*L(i)≥1*; *1≤i≤R*). We will compute *G(Q)*=the Grundy number for a linear board consisting of *Q* unmarked positions. *G(0)=0*. For *Q≥1*, we will consider all the *Q* positions *i* which the player can select: *G(Q)= mex({G(max{i-3, 0}) xor G(max{Q-i-2,0})) | 1≤i≤Q})*. The Grundy number of the initial board is *G(L(1)) xor … xor G(L(R))*. This way, we can compute a Grundy number for every instance of a game *i*. If *(x(i) mod 2=0)* then *GG(i)=0*; otherwise, *GG(i)*=the Grundy number of an instance of game *i*. The Grundy number of the entire combined game is *GG(1) xor … xor GG(K)*. If this number is *0*, then the second player has a winning strategy; otherwise, the first player has a winning strategy.

## 4. 2 AGENTS WITH DIFFERENT ROLES

### 4.1. GUESSING A SECRET STRING

We consider a secret string *S*, composed of symbols from the set *{0, 1, ..., K-1}*, and having an unknown length *L*. The player must ask questions in order to identify the string *S*. A question has the following form: *Ask(S')*, and the answer is *true*, if *S'* is a (not necessarily contiguous) subsequence of *S* (i.e. if *S'* can be obtained from *S* by deleting zero or more symbols), or *false*, otherwise. We want to determine the string *S* by asking as few questions as possible. I will present a strategy which asks at most *(K+1)·(L+1)* questions.

We will identify the string *S* one step at a time. We will maintain a representation *SR* of *S*, having the following structure. *SR* will be a sequence of zones, where each zone is of one of the following three types: *uncertain zone* (type *1*), *empty zone* (type *2*), and *certain zone* (type *3*). Before and after every uncertain or empty zone there is a certain zone (except, possibly, for the first and last zone of *SR*). A certain zone is composed of just one symbol. Let's assume that *SR* consists of *Q* zones (numbered, in order, from *1* to *Q*). The type of zone *i* is denoted by *ztype(i)* (*1≤i≤Q*). Initially, *SR* consists of only one uncertain zone. We will adjust the representation *SR* one step at a time, in rounds, until it will contain no more uncertain zones. When that happens, the concatenation of the symbols of the certain zones (from the first one to the last) will be the secret string *S*. At every round we will choose the first uncertain zone *i* from *SR* (the one with the lowest index). Let *i-1* and *i+1* be the certain zones before and after the uncertain zone *i* (if they exist). Let *cs(i-1)* and *cs(i+1)* be the symbols corresponding to these two zones (if *i-1=0* then *cs(i-1)=0*; if *i+1>Q* then *cs(i+1)=0*). Let *cstart=max{cs(i-1), cs(i+1)}*. We will consider, one at a time, every character *c* (*cstart≤c≤K-1*) and we will construct the string *S'(c)*, as follows: we concatenate all the symbols of the certain zones *j'<i* (from the lowest index to the largest one), then we add the symbol *c*, and then we add at the end the concatenation of all the symbols of the certain zones *j''>i* (from the lowest index to the largest one). Then, we ask the question *Ask(S'(c))*. When we get an affirmative answer, we break the loop (i.e. we do not construct the strings *S'(c')* with *c<c'≤K-1*) and then we modify the representation *SR*. The uncertain zone *i* will be replaced by an uncertain zone, followed by a certain zone containing the symbol *c*, and then followed by another uncertain zone. These *3* zones are inserted in *SR* in the place of the former uncertain zone *i*. If the answer is negative for every question, then we transform the uncertain zone *i* into an empty zone. The round ends either by replacing the uncertain zone *i* by three other zones, or by turning it into an empty zone. Then, we will move to the next round. The algorithm ends when *SR* contains no more uncertain zones. We notice that we ask at most *K* questions at every round. The total number of rounds is at most *2·L+1*, because: *(1)* there may be at most *L* rounds in which a new certain zone is created; *(2)* there will never be more than *L+1* uncertain zones which are turned into empty zones, during the execution of the algorithm. So, apparently, we may get to ask at most *2·K·(L+1)* questions. However, because at every round we do not consider the symbol *c* in the range *[0,K-1]*, but in the range *[cstart,K-1]*, the total number of questions is at most *(K+1)·(L+1)*.

### 4.2. HOTTER OR COLDER

In this section we consider the following resource discovery problem, modelled as a guessing game. An agent thinks of a secret natural number *S* from the interval *[1,N]* (the value of the resource amount). At the first question, a second agent (the player) asks an integer number *x* and expects no answer. At each of the next questions, whenever the player asks an integer number *y* (and at the previous question he/she asked a number *x*), it will receive the answer *Hotter* (*Colder*), depending on whether the secret number *S* is closer to (farther from) *y* than to (from) *x*. If *|S-x|=|S-y|*, any of the two answers may be received. The game ends when the player is absolutely sure which the secret number *S* is. We want to find a strategy which asks a minimum number of questions in the worst case (i.e. no matter what the secret number is). We consider two versions of this problem, one in which the number asked at every question must be a valid number (i.e. it must be one of the potential values for *S*, considering all the previous answers), and one in which it doesn't need to be a valid number. Both solutions are similar. We will compute a table *T(a,b,x)=* the minimum number of questions required to find *S* in the worst case, if *S* is within the interval *[a,b]* and the player asked *x* at the previous question. The answer will be *min{T(1,N,x)|1≤x≤N}* (at the first question, it makes no sense to ask for an invalid number). If *a=b*, then *T(a,b,x)=0*. For *a<b*, we will consider all the possibilities for the number *y* to be asked at the next question. In the first version, we will consider that *y* is between *a* and *b*; in the second version, *y*'s range is computed such that either *(x+y)/2* belongs to the interval *[a,b]*, or *y* is closer to *[a,b]* than *x* (we define *distance(z, [a,b])=if a≤z≤b then 0 else min{|z-a|, |z-b|}*). We now need to evaluate the maximum number of questions that the strategy will need to ask in the future, in case it asks *y* at the next question. If the answer is *Hotter*, then the secret number belongs to the intersection of *[a,b]* with *closerPart(y,x)*; if the answer is *Colder*, the secret number belongs to the intersection of *[a,b]* with *closerPart(x,y)*. *closerPart(u,v)* is defined as follows: *if (u≤v) then [u, floor((u+v)/2)] else [ceil((u+v)/2), u]*. Let *[c',d']* be the new (reduced) interval to which the secret number belongs if the answer is *Hotter*, and *[c'',d'']* the interval to which it belongs if the answer is *Colder*. The maximum number of questions which may need to be asked after asking *y* at the next question is *Q(a,b,x,y)=max{T(c',d',y), T(c'',d'',y)}*. Thus, for *a<b*, *T(a,b,x)=1+min{Q(a,b,x,y)| y obeys the conditions mentioned above}*. The time complexity of this solution is roughly $O(N^4)$, but can be reduced to $O(N^3)$, by using the following observation: *T(a,b,x)=T(1, b-a+1, x-a+1)*. This observation says that the actual interval *[a,b]* is not important for computing the number of required questions, only its length is. Thus, we will compute *T'(L,x)=*the minimum number of questions required to find *S* in the worst case, if *S* is within the interval *[a,b]=[1,L]* and the player asked *x* at the previous question (whenever the interval *[a,b]* is mapped to the interval *[1,L]*, *x* is also decreased by *(a-1)*, in order to maintain its position relative to *a* and *b*). With this definition, we have *Q'(L,x,y)=max{T'(d'-c'+1, y-c'+1), T'(d''-c''+1, y-c''+1)}* (instead of *Q(a,b,x,y)*) and *T'(L,x)=1+min{Q'(L,x,y)| y obeys the given conditions}*.

The problem can be extended as follows. Let's assume that before getting the first answer, the player must ask $D \geq 1$ questions. Then, at the $(D+1)^{th}$ question, the $(D+1)^{th}$ value is compared against the first value and we get the answer *Hotter* or *Colder* as before. Then, at every question $qu \geq D+1$, the value asked at that question is compared against the value asked at the question $qu$-$D$ and the answer *Hotter* or *Colder* is given. The case described so far is equivalent to the extended problem for $D=1$. The strategy can be adapted as follows. The index $x$ of the values $T(a,b,x)$ and $Q(a,b,x,y)$ is replaced by a tuple of $D$ values, representing the previous $D$ values asked: $x, x_2, ..., x_D$ (we denoted the first value by $x$, instead of $x_1$, on purpose). Then, $Q(a, b, x, x_1, ..., x_D, y) = \max\{T(c', d', x_2, ..., x_D, y), T(c'', d'', x_2, ..., x_D, y)\}$ and $T(a, b, x, x_1, ..., x_D) = 1+\min\{Q(a, b, x, x_1, ..., x_D, y) | y$ obeys the specified conditions$\}$. For the case in which the indices $a$ and $b$ are replaced by the length $b$-$a$+1 of the interval $[a,b]$, we perform the same substitutions. Then, we have $Q'(L, x, x_2, ..., x_D, y) = \max\{T'(d'-c'+1, x_2-c'+1, ..., x_D-c'+1, y-c'+1), T'(d''-c''+1, x_2-c''+1, ..., x_D-c''+1, y-c''+1)\}$ and $T'(L, x, x_2, ..., x_D) = 1+\min\{Q'(L, x, x_2, ..., x_D, y) | y$ obeys the specified conditions$\}$. The time complexity now becomes $O(N^{D+3})$ (for the first solution), or $O(N^{D+2})$ (for the second solution).

### 4.3. FINDING A COUNTERFEIT COIN

We are given a set of $n \geq 3$ coins, out of which one is different (heavier or lighter) than the others. We have a balance with *2* pans. We can place any equal number of coins $k$ ($1 \leq k \leq n/2$) on each pan and compare the total weight of the coins on the left pan to the total weight of the coins on the right pan (i.e. ask a question). There are *3* possible outcomes: the coins on the left pan are lighter than, heavier than, or of the same weight as those on the right pan. We assign to each coin $i$ a set $C(i)$ which consists of all the possible types coin $i$ may have (e.g. *normal, lighter,* or *heavier*). Initially, all the sets $C(i)$ are equal to *{lighter, heavier, normal}*. The current state $S$ of the game consists of all the sets $C(i)$. After every comparison performed, some of the sets $C(i)$ will be reduced. We define the uncertainty $U(S)$ of a state $S$ to be equal to *-1*, plus the sum of the values $(|C(i)|-1)$ ($1 \leq i \leq n$); $|C(i)|$ denotes the cardinality of the set $C(i)$. At each moment during the game (when the uncertainty is not zero), every set $C(i)$ (coin $i$) can be of *4* types: *(1) {normal, lighter, heavier}*; *(2) {normal, lighter}*; *(3) {normal, heavier}*; *(4) {normal}*. We denote by $num(i)$ the number of sets (coins) of type $i$ ($1 \leq i \leq 4$). Obviously, $num(1)+num(2)+num(3)+num(4)=n$. When asking a question, we can place any combination of coins of each type on each pan, with the condition that coins of type *4* are placed on at most one of the pans. If the result of a question $Q$ is that the coins on the left (right) pan are lighter than those on the right (left) pan, then we remove the element *heavier* from the sets $C(i)$ of the coins $i$ on the left (right) pan and the element *lighter* from the sets $C(j)$ of the coins $j$ on the right (left) pan. Moreover, the sets $C(k)$ of all the coins $k$ which were not placed on any pan are set to *{normal}*. If the sets of coins on the two pans have equal weights, then we set the sets $C(i)$ of the coins $i$ placed on any of the two pans to *{normal}*. If, at some point, only one set $C(i)$ is different than *{normal}* and contains only two elements, then coin $i$ is the different coin: if (*lighter*$\in C(i)$), then coin $i$ is lighter than the other $n$-1 coins; otherwise, coin $i$ is heavier.

Notice that before receiving the first answer in which the coins on one of the pans are lighter or heavier than those on the other pans, we only have sets of two types: *1* and *4*. After receiving the first answer where the coins on the two pans have different weights, then we have no more sets of type *1*. As long as we only have sets of types *1* and *4*, we have $O(n^2)$ possible questions (at each step). A question is uniquely defined by the total number of coins $k$ on each of the two pans and by the number $x \leq \min\{num(1), k\}$ of coins of type *1* which are placed on the left pan. The right pan will contain $k$ coins of type *1* and the left pan will contain $k$-$x$ extra coins of type *4* (we must have $x+k \leq num(1)$). If the weight of the two pans is equal, then the total uncertainty decreases by $2 \cdot (k+x)$. If one of the pans is lighter (heavier) than the other, then the total uncertainty decrease by $k+x+2 \cdot (num(1)-k-x) = 2 \cdot num(1)-k-x$. If we denote by $S(k,x)=k+x$, we notice that in the first case the total uncertainty decreases by $2 \cdot S(k,x)$, and in the second case, by $2 \cdot num(1)-S(k,x)$. The best case is when the minimum value of the two uncertainty decrements is as large as possible. This occurs when $2 \cdot S(k,x)=2 \cdot num(1)-S(k,x)$ and, thus, $S(k,x)$ is equal either to $\lfloor 2/3 \cdot num(1) \rfloor$ or to $\lceil 2/3 \cdot num(1) \rceil$. Thus, we have only $2=O(1)$ possibilities for choosing the next question. Once $S(k,x)$ is fixed, we can choose $k=\lceil S(k,x)/2 \rceil$ and $x=S(k,x)-k$.

When we have only coins of types *2*, *3* and *4*, there are $O(n^4)$ possible questions for the next step. A question is defined by the number of coins $k$ placed on each pan, the number $x$ of coins of type *2* ($x \leq num(2)$) and the number $y$ of coins of type *3* placed on the left pan ($y \leq num(3)$ and $x+y \leq k$), and the number $z$ of coins of type *2* placed on the right pan. The left pan will also contain $k$-$(x+y)$ coins of type *4* and the right pan will also contain $k$-$z$ coins of type *3*. We must have $num(2)$-$x \geq z$ and $num(3)$-$y \geq k$-$z$, i.e. $num(2)$-$x \geq z \geq k+y$-$num(3)$ ($x+y+k \leq num(2)+num(3)$). If the result of such a question is that the coins on both pans have equal weights, then the total uncertainty decreases by $(x+y+z+k-z)=(x+y+k)$. If the coins on the left pan are lighter, then the uncertainty decreases by $(y+z+num(2)-x-z+num(3)-y-(k-z)) = (num(2)+num(3)-x+z-k)$. If the coins on the right pan are lighter, then the uncertainty decreases by $(k-z+x+num(2)-x-z+num(3)-y-(k-$

$z$))=($num(2)+num(3)-z-y$). Let's assume that $x+y=S(x,y)$, $y+z=S(y,z)$ and $D(x,z)=x-z$. Obviously, the uncertainty depends on the values $S(x,y)$, $S(y,z)$ and $D(x,z)$, rather than on the actual $x$, $y$ and $z$ values. Let's consider a different value $z'$, such that $z-z'=dz$. Then we have $y'=y+dz$, $x'=x-dz$ and $x'-z'=x-dz-(z-dz)=x-z$. Thus, for any value of $z$, we can find a pair $(x,y)$ which maintains the same values of $S(x,y)$, $S(y,z)$ and $D(x,z)$. The maximum possible difference $dzmax$ (by which a chosen value of $z$ can be reduced) has the following properties: $z-dzmax \geq 0$, $x-dzmax \geq 0$, $y+dzmax \leq num(3)$ and $y'+k-z' \leq num(3)$ $\Leftrightarrow$ $y+dzmax+k-(z-dzmax) \leq num(3)$. From these constraints it is trivial to compute the largest possible value of $dzmax$, given the values of $x$, $y$ and $z$. The key observation here is that if $x$, $y$ and $k$ are fixed, then $z$ can be chosen to be as small as possible (but without violating the constraints). Choosing $z$ can be done in $O(1)$ time, leaving us with only $O(n^3)$ possibilities, for the parameters $x$, $y$ and $k$.

A better approach is the following. We would like to maximize the minimum value of the uncertainty decrements of each of the three situations. We can do this by binary searching this minimum (integer) value $W$. Then, we have the following constraints: $x+y+k \geq W$, $num(2)+num(3)-x+z-k \geq W$, $num(2)+num(3)-z-y \geq W$, $x \geq 0$, $y \geq 0$, $z \geq 0$, $x+z \leq num(2)$, $y+k-z \leq num(3)$, $x+y \leq k$, $z \leq k$. These equations define half-hyperspaces in the hyper-space with four dimensions, in which every dimension corresponds to one of the parameters $x$, $y$, $z$, and $k$. $W$ is a feasible value (i.e. all the constraints can be satisfied), if the intersection of all these half-hyperspaces is non-empty and contains at least one point with integer coordinates (inside of it, or on one of its sides). If $W$ is feasible, then we will test a larger value of $W$ in the binary search; otherwise, we will test a smaller value. Since we have $O(1)$ half-hyperspaces, we can compute their intersection in $O(1)$ time, obtaining a convex 4-dimensional polyhedron. However, we are not aware of any efficient method of checking if the polyhedron contains any point with integer coordinates. If we consider every possible value of $k$, then, for every such fixed value of $k$, the only free parameters will be $x$, $y$, and $z$. In this case, the specified constraints define half-spaces in a 3D space (with the dimensions corresponding to the parameters $x$, $y$ and $z$). The intersection of the half-spaces is now a 3D convex polyhedron, but we still don't know how to find a point with integer coordinates inside of it. If we fix the values of two parameters, e.g. $k$ and $x$ (i.e. we consider every possible value of $k$ and $x$ and then find some suitable values for the parameters $y$ and $z$), then the specified constraints become half-planes delimited by lines in the plane (the 2 dimensions of the plane correspond to the two free parameters, e.g. $y$ and $z$). These lines have only 4 different orientations: parallel to the horizontal or vertical axes, or parallel to the two axes rotated by 45 degrees counter-clockwise. If the intersection of the half-planes is non-empty, then the resulting polygon is a convex polygon with $O(1)$ (at most 6) sides. The polygon's vertex coordinates are either integers or are of the form $(p+q)/2$, where $p$ and $q$ are integer numbers. However, a simple analysis shows us that, if the intersection is non-empty, then at least one of the vertices of the polygon must have integer coordinates. Thus, it is sufficient to check if the intersection is non-empty and, afterwards, look at the vertices of the intersection. This way, we only need to test $O(n^2)$ possibilities (e.g. for the values of the parameters $k$ and $x$). The intersection of the original 4D half-hyperplanes is non-empty and contains a point with integer coordinates if there exists at least one pair of values $(k,x)$ for which we can find a point with integer coordinates in the corresponding intersection polygon (or if there exists at least a value of $k$ for which the intersection of the obtained 3D half-spaces is non-empty and contains at least a point with integer coordinates).

We conjecture that a property similar to the one for the 2D case also holds for the 4D case. Thus, we only need to check if the intersection of the half-hyperspaces is non-empty and then only look at the contour of the intersection polyhedron (e.g. its vertices, edges, or faces) in order to find a point with all of its coordinates integer numbers. With this conjecture, we are able to decide the next question in $O(log(n))$ time (for the second case), instead of $O(n^2 \cdot log(n))$, and in $O(1)$ time for the first case. In both cases, at each step, we set the values of all the relevant parameters for each of the considered possibilities and we choose that possibility for which the worst case uncertainty decrement (i.e. the minimum possible uncertainty decrement) is as large as possible; the chosen possibility will be the question asked next.

### 4.4. MAXIMIZING WORST-CASE BET REVENUES

Let's consider the following game. A player has $X$ monetary units initially. A box contains $N+R$ objects: $N$ black objects and $R$ red objects. The player cannot see the objects inside the box. The game is played for $N+R$ rounds. At every round, the player bets any percent $p$ between 0 and 100% of its current sum on one of the two colors: *black* or *red*. Then, an object from the box is extracted (and never placed back). If the color of the extracted object is the same as the color on which the player bet, then the player gains an amount of monetary units equal to its bet; otherwise, the player loses the sum it bet at the current round. We would like to maximize the final amount of monetary units in the worst-case. In order to do this, we need to compute a strategy which tells the player what percent $p$ to bet at every round and on which color. We will compute $pmax(i,j)$=the maximum multiplication factor by which the player's initial amount can be multiplied in the end, if the box initially contains *i* black objects and *j* red objects. To be more precise, if the player initially has *X* monetary units, then there is a strategy which guarantees him/her at least *pmax(i,j)·X* monetary units in the end, and no other strategy can guarantee more than that. We notice that $pmax(0,j)=pmax(j,0)=2^j$ (because, at every round, the player can bet its entire amount of monetary units). For $i\geq 1$ and $j\geq 1$, let's assume that the player bets a percent *p*. on the color *black*. If it is right, then its sum will increase *(1+p)·pmax(i-1, j)* times; if it is wrong, then its sum increases *(1-p)·pmax(i, j-1)* times. The percent *p* must be chosen such that, in the worst case, its final sum is as large as possible. This occurs when *(1+p)·pmax(i-1, j)=(1-p)·pmax(i, j-1) => p=(pmax(i,j-1) - pmax(i-1,j)) / (pmax(i-1, j) + pmax(i, j-1))*. If $p\geq 0$, then *pmax(i,j)=(1+p)·pmax(i-1, j)*. If *p<0*, then the player should not bet on the *black* color, but on the *red* color instead. By using the same argument we obtain *(1+p)·pmax(i, j-1)=(1-p)·pmax(i-1, j) => p=(pmax(i-1,j) – pmax(i,j-1)) / (pmax(i-1, j) + pmax(i, j-1))* and *pmax(i,j)=(1+p)·pmax(i, j-1)*. The maximum final amount of monetary units which can be guaranteed by the best strategy is *X·pmax(N,R)*. The time complexity of the described algorithm is *O(N·R)*.

## 4.5. ALGEBRAIC COMPUTATIONS

We consider *M* triples of numbers: *P*, *Q* and *N* and we know that *P=a+b* and *Q=a·b* (for some numbers *a* and *b*). We need an efficient algorithm which computes, for every triple, the values $a^N+b^N$. A first solution, with *O(N)* time complexity, is the following. We define $SP(pow)=a^{pow}+b^{pow}$. *SP(0)=2* and *SP(1)=P*. For $2\leq pow\leq N$, we have *SP(pow)=SP(pow-1)·P-SP(pow-2)·Q*. More precisely, we have $(a^{pow-1}+b^{pow-1})\cdot(a+b)-(a^{pow-2}+b^{pow-2})\cdot a\cdot b = a^{pow}+b^{pow}+b\cdot a^{pow-1}+a\cdot b^{pow-1}-b\cdot a^{pow-1}-a\cdot b^{pow-1}=a^{pow}+b^{pow}$. A faster solution is based on using the characteristic polynomial of the recurrence relation mentioned above. We compute $delta=sqrt(P^2-4\cdot Q)$; $c_1=(P+delta)/2$; $c_2=(P-delta)/2$ (*sqrt(x)* denotes the square root of *x*). Let $V=(P-2\cdot c_1)/(c_2-c_1)$ and *U=2-V*. We will then determine the binary representation of the number *N*: *b(BMAX), b(BMAX-1), ..., b(0)* (*N*=the sum of the values $b(j)\cdot 2^j$ for $0\leq j\leq BMAX$); we have *b(BMAX)=1* (the most significant bit of *1* in the binary representation of *N*). We will compute $d_1=c_1^N$ and $d_2=c_2^N$ in *O(log(N))* steps, by using this binary representation: we initialize $d_1=d_2=1$ and then we traverse the bits *j* from *j=BMAX* down to *j=0*; for every bit *j*, we set $d_1=d_1^2$ and $d_2=d_2^2$; then, if *b(j)=1*, we set $d_1=d_1\cdot c_1$ and $d_2=d_2\cdot c_2$. The final answer is $U\cdot d_1+V\cdot d_2$. Note, however, that *V=1* and, thus, *U=1*, too.

# 5. AGENT PURSUING GAMES ON GRAPHS

## 5.1. PURSUING A SET OF ROBBERS

In a directed graph with *n* vertices and *m* edges there are *A* cop agents (cops) and *B* robber agents (robbers): every cop *i* is initially located at vertex *P(i)* ($1\leq i\leq A$) and every robber *j* is initially located at vertex *S(j)* ($1\leq j\leq B$). We are also given a sequence of *K* pairs *(type(0), idx(0)), ..., (type(K-1), idx(K-1))*, meaning that at time moment *T* ($T\geq 0$) the agent that will perform the move is of type *type(T mod K)* (i.e. cop or robber) and its index is *idx(T mod K)* (between *1* and *A* for cops, and between *1* and *B* for robbers). A move consists of moving the agent from its current vertex *i* to an adjacent vertex *j* such that the directed edge *i->j* exists in the graph (the graph is allowed to have loops, i.e. *j=i*). The cops win if at least *B'≤B* robbers are captured (a robber is captured whenever a cop moves to the same vertex as the robber). The robbers win if at least *B''≤B* robbers reach their safe vertices: every robber *j* has a set of vertices *H(j)* representing its safe vertices – if it reaches one of these vertices, it cannot be captured by any cop anymore. If the game continues indefinitely, then it ends as a draw. A state of the game consists of a tuple with *A+B+1* values: *(pozc(1), ..., pozc(A), pozr(1), ..., pozr(B), p)*; *p* is the index of the current move ($0\leq p\leq K-1$). *pozc(i)* is the vertex where the $i^{th}$ cop is located ($1\leq i\leq A$) and *pozr(j)* is the vertex where the $j^{th}$ robber is located ($1\leq j\leq B$). We also consider two special extra positions for the robbers, which indicate if the robber was captured, or if it already arrived to one of its safe places. Thus, based on this representation, we can decide for some of the states if they lead to the victory or defeat of the cops (robbers) : for instance, those states where at least *B'* robbers are captured are winning states for the cops, while those with at least *B''* robbers in their safe places are winning states for the robbers. More generally, some states of the game are known to lead to the victory or defeat of the cops (robbers) or to a draw.

We will construct the state graph *GS*, by adding directed edges from every state *(pozc(1), ..., pozc(A), pozr(1), ..., pozr(B), q)* to every state *(pozc'(1), ..., pozc'(A), pozr'(1), ..., pozr'(B), (q+1) mod K)*, where only the position of the agent (cop or robber) whose turn is to move next is changed, and all the other positions remain the same. If *GS* is acyclic, then we can easily use any of the algorithms described in Section *2* for deciding the outcome. Otherwise, we can use the other techniques presented in Section *2*: we can either introduce an extra parameter $T\geq 0$, indicating the index of the current move, which is bounded from above by *V(GS)+1* (in which case we drop the index *q* from the game state ; *q* can be easily computed as *q=T mod K*), or we can use the iterative solution. Since $V(GS)=O(K\cdot N^{A+B})$, the time complexity of any of the presented approaches is high.

## 5.2. THE CASE WITH 1 COP AND 1 ROBBER

We now consider the same game as in the previous subsection, with the following restrictions. There is only one cop and one robber and they move alternately. The game starts when the cop chooses an initial vertex, after which the robber chooses an initial vertex. Only after the initial choices, the cop and the robber start moving alternately. Moreover, the graph where all the action takes place is undirected. The robber has no safe place. Thus, the cop wins if it moves to the same vertex as the robber, while the game ends as a draw (or, equivalently, the robber wins) if the robber can escape the cop indefinitely. At first, we should notice that we can use the methods presented in the previous subsection. However, this problem was considered in [13] and the following generic algorithm was given for deciding if the cop has a winning strategy. We say that a vertex $X$ *dominates* another vertex $Y$ if the edge $(X,Y)$ exists in the graph and for every other vertex $Z$, such that $Z$ is a neighbor of $Y$, $Z$ is also a neighbor of $X$ (vertex $X$ may also have other neighbors except $Y$ and $Y$'s neighbors). A vertex $Y$ is *dominated* if there exists at least one vertex $X$ such that $X$ dominates $Y$. It should be obvious that, if the cop has a winning strategy, before the last move of the robber the cop is located at a vertex $X$ and the robber is located at a vertex $Y$, such that $X$ dominates $Y$. If the graph contains no dominated vertex $Y$, then the cop has no winning strategy (the robber will be able to escape the cop indefinitely). The following observation is paramount. If a graph $G$ contains a dominated vertex $Y$, then the cop has a winning strategy in $G$ if and only if it has a winning strategy in $G'=G\setminus Y$ (i.e. the graph $G$ from which we remove the vertex $Y$, together with all of its adjacent edges). With this observation, we have the following algorithm:

**1.** while the graph has at least *2* vertices and contains at least one dominated vertex, then find any dominated vertex $Y$ and remove it from the graph

**2.** if the graph has only one vertex left, then the cop has a winning strategy; otherwise, the robber will be able to escape the cop indefinitely.

Step *1* is executed $O(n)$ times. Thus, the essential part of the algorithm is finding a dominated vertex $Y$ efficiently. The naive solution is to consider every vertex $Y$ at every iteration of Step *1*. Then, for every such vertex $Y$, we consider every vertex $X$ which is a neighbor of $Y$, and then we check if every other neighbor $Z$ of $Y$ ($Z \neq X$) is also a neighbor of $X$ (the check can easily be performed if we represent the graph by using its adjacency matrix). If all the conditions hold, we found a dominated vertex $Y$ and we do not consider any other vertex until the next iteration of Step *1*. Afterwards, we remove $Y$ from the graph (e.g. by marking it as removed, and by removing the edges between $Y$ and its neighbors from the adjacency matrix). This approach has $O(n^3)$ time complexity per iteration and, thus, $O(n^4)$ overall. However, in practical settings, this naive solution is quite good, because: *(1)* in dense graphs (i.e. with many edges), a dominated vertex $Y$ is found quickly (if one exists); *(2)* in sparse graphs, the time complexity is lower than $O(n^3)$ per iteration.

Nevertheless, a smarter solution exists. Initially, we will compute all the values $NVC(X,Y)$=the number of common neighbors between the vertices $X$ and $Y$. We do this in $O(n^3)$ time by initializing $NVC(*,*)=0$ and then considering every vertex $Z$: for every vertex $Z$ we consider every pair of neighbors $X$ and $Y$ of the vertex $Z$ and we increment $NVC(X,Y)$ and $NVC(Y,X)$ by *1*. Moreover, we will also compute the values $NV(X)$=the number of neighbors of the vertex $X$, for every vertex $X$ of the graph. Then, at every iteration of Step *1*, we will consider every vertex $Y$ and check if it is dominated. We do this in $O(n)$ time, by considering every neighbor $X$ of $Y$ and checking if $NVC(X,Y)=NV(Y)-1$. If the condition holds, then vertex $Y$ is dominated by vertex $X$. In order to remove vertex $Y$ from the graph we first consider every neighbor $Z$ of $Y$ and decrease $NV(Z)$ by *1* (and also remove $Y$ from the list of neighbors of vertex $Z$). Then, we consider every pair $(Z,X)$ of neighbors of vertex $Y$, and we decrement $NVC(Z,X)$ and $NVC(X,Z)$ by *1*. As we can see, the time complexity per iteration is $O(n^2)$. Thus, the overall time complexity is $O(n^3)$.

## 6. EQUITABLE RESOURCE ALLOCATION

### 6.1. UNCONSTRAINED REALLOCATIONS

We consider the following problem. We have $n$ resource containers, numbered from *1* to *n*. Each container $i$ contains an amount of resources $r(i) \geq 0$. We want to perform reallocations such that, in the end, every container contains the same amount of resources $q=(r(1)+...+r(n))/n$. A reallocation consists of taking any amount $x$ of resources from a container $i$ and moving them to any other container $j$. We do not care about minimizing the total number of reallocations, but this number should be of the order $O(n)$. We will sort the containers in increasing order of their resource amounts: $r(p(1)) \leq ... \leq r(p(n))$. We initialize a variable *left=1* and a variable *right=N*. While *left<right* we perform the following actions: if $r(p(left))=q$, then *left=left+1*; else, if $r(p(right))=q$, then *right=right-1*; otherwise: *(1)* $x=min\{q-r(p(left)), r(p(right))-q\}$; *(2)* we move $x$ resource units from the container $p(right)$ to the container $p(left)$, i.e. we set $r(p(right))=r(p(right))-x$ and $r(p(left))=r(p(left))+x$. In the end, all the containers will contain *q* resource units. The time complexity of the algorithm is *O(n·log(n))*, or *O(n)* if the containers are given sorted according to their resource amounts.

### 6.2. MAXIMIZING THE AMOUNT OF RESOURCES

Along a line there are *n* containers, numbered from *1* to *n* (form left to right). Every container *i* contains an amount *r(i)* of resources. We want to redistribute these resources in such a way that the minimum amount of resources in any container is maximized. In order to perform the redistribution, the resources can be transported along the line. If we transport resources on the line segment between the containers *i* and *i+1* ($1 \leq i \leq n-1$), *q(i)* resources are consumed because of the transportation. We will binary search the maximum value *Xopt* such that there is a redistribution strategy which leaves in every container at least *Xopt* resource units. Let's assume that we selected a value *X* during the binary search. We will now perform a feasibility test. If *X* is feasible, then we can obtain at least *X* resource units in every container and, thus, we will consider larger values of *X* in the binary search next; if *X* is not feasible, then we will consider smaller values during the binary search next. The feasibility test has a linear (i.e. *O(n)*) time complexity. We will traverse the containers from *1* to *n* (in increasing order) and we will maintain a variable *E*, representing the surplus (if *E≥0*) or the uncovered required amount (if *E<0*) of resources. We start with *E=0*. Let's assume that we reached the container *i*. If *r(i)≥X*, then we increment *E* by *(r(i)-X)*; otherwise, if *r(i)<X*, then we decrement *E* by *(X-r(i))* (thus, in both cases, we can increment *E* by *(r(i)-X)*). If *i<n* then, before going to the next container *i+1*, we perform the following actions: *(1)* if *E≥0* then we set *E=max{E-q(i), 0}* ; *(2)* if *E<0* then we set *E=E-q(i)*. After traversing all the containers, if *E≥0* then *X* is a feasible value; otherwise, *X* is not feasible. The problem can be generalized by associating to every container *i* a non-decreasing function *$f_i(X)$* and we want to find the largest value of *X* such that every container *i* contains at least *$f_i(X)$* resources. The feasibility test is modified slightly: when we reach a container *i* during their traversal, we increment *E* by *(r(i)-$f_i(X)$)* instead of *(r(i)-X)*.

### 6.3. RESOURCE REALLOCATIONS IN A TREE

We consider a tree with *n* vertices (numbered from *1* to *n*). Every vertex *i* contains an integer number *b(i)≥0* of resource units. We can perform reallocations which consist of moving one resource unit from a vertex *i* to a neighboring vertex *j*; the cost of such a move is *c(i,j)* monetary units (*c(i,j)* may be different than *c(j,i)*) We want to compute a strategy with minimum total cost such that, in the end, every vertex *i* contains exactly *q(i)* resource units (the sum of the *b(i)* values is equal to the sum of the *q(i)* values; $1 \leq i \leq n$). We will root the tree at a vertex *r* and we will establish parent-son relationships, based on the chosen root. Then, we will traverse the tree from the leaves towards the root. For every node *i* we will compute *S(i)*=the surplus of resources from vertex *i*'s subtree (*S(i)* may be negative). We will maintain a variable *C*, representing the total cost of the moves which we need to execute (*C* is zero, initially). Let's assume that we have to handle a leaf *i* of the tree. If *b(i)<q(i)*, then *S(i)=(b(i)-q(i))* and we increment *C* by *|S(i)|·c(parent(i),i)* (*q(i)-b(i)* resource units will have to be moved from leaf *i*'s parent to leaf *i*); if *b(i)≥q(i)* then *S(i)=b(i)-q(i)* and we increment C by *S(i)·c(i,parent(i))* (we will have to move *b(i)-q(i)* resource units from the leaf *i* to its parent). Let's assume that we reached an internal node *i*. We initialize *S(i)* to the sum of the values *S(j)* of the sons *j* of the vertex *i*. If *b(i)<q(i)*, we will decrement *S(i)* by *(q(i)-b(i))* (*q(i)-b(i)* resource units must be brought to vertex *i*); if *b(i)≥q(i)*, we will increment *S(i)* by *(b(i)-q(i))* (*b(i)-q(i)* resource units will have to be moved from vertex *i* towards other vertices, either in vertex *i*'s subtree, or towards vertex *i*'s parent if *i≠r*). If *S(i)<0*, then *|S(i)|* resource units will have to be moved from vertex *i*'s parent to vertex *i*; thus, we increment *C* by *|S(i)|·c(parent(i),i)*. If *S(i)≥0* then we will have to move *S(i)* resource units from vertex *i* towards vertex *i*'s parent; thus, we increment *C* by *S(i)·c(i,parent(i))*. Every time, we only performed those moves which were strictly required; some moves were performed before having enough resources in the vertex from which the resources had to be moved (for the cases with negative *S(*)*), but these resources will be brought later by the algorithm. *C* will be the minimum total cost of the moves which need to be performed such that, in the end, every node *i* contains exactly *q(i)* resource units.

## 7. RELATED WORK

Two-player games with identical player roles have been studied extensively in the literature and many techniques for computing optimal strategies were developed. We refer the reader, for instance, to [7] and [9]. Two-player games with different roles have also been studied from multiple perspectives; see [2] for a problem similar to the one discussed in subsection 4.2. An excellent survey of single- and two-player games which can be solved by dynamic programming is presented in [15]. An interesting hidden evader pursuit problem was brought to my attention by C. Negruseri. A hidden evader is located at an unknown vertex of a graph *G*. At each time step, the searcher performs a move and then the evader performs a move. When the searcher peforms a move, it selects a vertex of the graph and checks if the evader is located at that vertex. If the evader is there, then the evader is captured; otherwise, the searcher gathers the knowledge that the evader is not located at that vertex. The evader's move consists of moving from its current vertex $v$ to a vertex $u$ which is adjacent to $v$ (i.e. it is connected to $v$ by an edge); the evader cannot remain in the same vertex for two consecutive time steps. We want to know if the searcher has a strategy according to which it will eventually find the evader, no matter where the evader was located initially. The problem actually asks to characterize the graphs on which such a strategy exists. It turns out that such a strategy exists only on graphs in which every connected component is an *extended* caterpillar. An extended caterpillar is a tree graph, in which we can identify a central path of vertices. Then, every vertex $v$ on that path may have as neighbors any number of leaves (vertices of degree $1$) and any number of *leaf-neighbors* (vertices whose neighbors are only leaves, plus the vertex $v$), except for the (at most) two neighbors on the central path. Verifying if a connected component is an extended caterpillar is easy. First, we check if it is a tree. Then, we mark all the leaves. Afterwards, we mark all the leaf-neighbors (i.e. all the vertices having only one neighbor which is not a leaf). The remaining unmarked vertices must be located on the central path, i.e. every such vertex must have at most $2$ neighbors which are not leaves or leaf-neighbors. Checking if a tree is a caterpillar is performed similarly: every vertex with degree greater than $1$ is allowed to have at most $2$ neighbors with degrees greater than $1$. Thus, linear time recognitions algorithms for caterpillars and extended caterpillars exist. We should note that a caterpillar is also equivalent to a tree interval graph (i.e. the class of interval graphs which are also trees is equal to the class of caterpillar graphs).

Resource (re)allocation methods and strategies were discussed, for instance, in [3], [4], [6] and [8]. Assigning and computing resource costs is also an important issue, which was partly discussed in [11]. The counterfeit coin problem and variations of it have been studied extensively, for instance, in [1], [12] and [14]. Some cops-and-robbers games were discussed in [10] and [13].

## 8. CONCLUSIONS AND FUTURE WORK

In this paper we considered several types of rules for modelling the interaction between pairs of agents with contrasting interests, having identical or different roles. We expressed these rules in the context of several two-player games, for which we presented algorithmic techniques for optimizing the decision process of the involved agents. As future work, we intend to consider other types of rules for modelling the interactions between agents, like, for instance, rules based on resource negotiations, auctions, and several others.